\documentclass[final,3p,times,twocolumn]{elsarticle}

\usepackage{graphicx}
\usepackage{amssymb}
\usepackage{amsthm}
\usepackage{slashbox}
\newcommand{\la}{$\Lambda$}
\newcommand{\hla}{$\mathrm{^{3}_{\Lambda}H}$}
\newcommand{\he}{$\mathrm{^{3}He}$}
\newcommand{\hlafrac}{$\mathrm{^3_\Lambda H} / ( \mathrm{^3He} \times \frac{\Lambda}{p} )$}
\newcommand{\cbs}{$C_{\rm BS}$}
\newcommand{\sNN}{$\sqrt{s_{_\mathrm{NN}}}$ }
\journal{Physics Letters B}

\begin{document}

\begin{frontmatter}


\title{Searching for onset of deconfinement via hypernuclei and baryon-strangeness correlations
}

\author{S. Zhang$^{a}$}
\author{J. H. Chen$^{b}$\corref{cor1}}
\author{H. Crawford$^{c}$}
\author{D. Keane$^{b}$}
\author{Y. G. Ma$^{a}$}
\author{Z. B. Xu$^{d,e}$}
\cortext[cor1]{email address: jhchen@rcf.rhic.bnl.gov (J.H. Chen)}
\address{$^a$Shanghai Institute of Applied Physics, CAS, Shanghai, 201800, China}
\address{$^b$Kent State University, Kent, Ohio, 44242, USA}
\address{$^c$University of California, Berkeley, California, 94720, USA}
\address{$^d$Brookhaven National Laboratory, Upton, New York, 11973, USA}
\address{$^e$University of Science and Technology of China, Hefei, Anhui, 230026, China}

\begin{abstract}
We argue that the ratio $S_3 =$ \hlafrac~is a good representation
of the local correlation between baryon number and strangeness,
and therefore is a valuable tool to probe the nature of the dense
matter created in high energy heavy-ion collision: quark gluon
plasma or hadron gas. A multiphase transport model (AMPT) plus a
dynamical coalescence model is used to elucidate our arguments. We
find that AMPT with string melting predicts an increase of $S_3$
with increasing beam energy, and is consistent with experimental
data, while AMPT with only hadronic scattering results in a low
$S_3$ throughout the energy range from AGS to RHIC, and fails to
describe the experimental data.

\end{abstract}


\end{frontmatter}


Data from the Relativistic Heavy Ion Collider (RHIC) at Brookhaven
National Lab show evidence for partonic collectivity and other
likely signatures of quark gluon plasma (QGP) formation during the
early stages of the
collisions~\cite{BRAHMS-White-Paper,PHENIX-White-Paper,STAR-White-Paper,PHOBOS-White-Paper}.
Nevertheless, several important questions remain unresolved, such
as the beam energy where the QGP signatures first appear, and
other details of the transition between hadronic and deconfined
matter. Quantum Chromodynamics (QCD) predicts a critical point
separating a first-order phase transition and a smooth crossover
in the phase diagram of the hot and dense QCD
matter~\cite{LQCD-0,LQCD-1}. It is believed that large
fluctuations in phase space population or large correlation length
will be one of the experimental signatures of the QCD critical
point. Investigations of all of these questions began at the SPS
and the upcoming Beam Energy Scan at RHIC~\cite{STAR-BES-Long}
will provide an opportunity to study them in more detail.
Regardless of how difficult or easy it will be to uncover specific
experimental evidence of a critical point, it is a high priority
to identify and understand all the observables that offer a
prospect of discriminating between hadronic and deconfined matter
with good sensitivity.

In calculations from lattice QCD at high temperature, and in
models with an ideal quark gas or hadron resonance gas, the cross
correlations among the conserved charges show sensitivity to the
confined hadron phase or deconfined quark-gluon
phase~\cite{STAR-BES-Long,Cbs_PRL,Cbs_PRC,LQCD-2,LQCD-3}.
Specifically, in lattice QCD~\cite{LQCD-3}, the ratio
$\chi_{11}^{BS}$/$\chi_{2}^{B}$, the strangeness-baryon
correlation ($\chi_{11}^{BS}$) normalized by the baryon-baryon
correlation ($\chi_{2}^{B}$), approaches unity at high temperature
in a deconfined phase, and reaches 0.4 at low temperature in a
hadronic phase. The baryon-strangeness correlation coefficient
\cbs~was argued to be a robust observable to characterize the
nature of the system created in high energy heavy-ion collisions:
ideal QGP or strongly coupled QGP or hadronic
matter~\cite{Cbs_PRL,Cbs_PRC}. Although the local
baryon-strangeness correlation is a sensitive probe of the
partonic and hadronic phases as predicted by the Lattice
QCD~\cite{LQCD-2}, the proposed experimental observable (\cbs) is
based on global extensive quantities~\cite{Cbs_PRL,Cbs_PRC}.
Because it requires a measurement of the global baryon number and
strangeness in each event, an experimental analysis based on
\cbs~represents a considerable technical challenge. Further
detailed theoretical investigation indicated that a
recombination-like hadronization process and hadronic rescattering
both have the effect of blurring the fluctuation
signal~\cite{Cbs_recombination,Cbs_AMPT}.

On the other hand, hypernuclei are clusters of nucleons and
\la~hyperons~\cite{hypernuclei:1953}. The production of
hypernuclei happens through a coalescence mechanism by the
overlapping of the wave functions of protons, neutrons and
hyperons at the final stage of the collisions~\cite{Gutbrod:1976}.
This provides a local correlation of baryons and strangeness on an
event-by-event basis~\cite{Steinheimer:2008hr}. Specifically, the
deuteron yield is proportional to the baryon density while triton
($t$) and helium (\he) are a measure of baryon
correlation~\cite{Sato-Coalescence1981,Wang:1999xi}. Similarly,
hypertriton production is related to the primordial \la-p phase
space correlation. The ratio $S_3 =$ \hlafrac, which we call the
Strangeness Population Factor, shows model-dependent evidence of
sensitivity to the local correlation strength between baryon
number and strangeness, and is demonstrated in this paper to be a
promising tool to study the onset of deconfinement. The ratio
$S_3$ is quantitatively a good representation of
$\chi_{11}^{BS}$/$\chi_{2}^{B}$~\cite{LQCD-3} since $S_3$ contains
the local strangeness-baryon correlation in the numerator and the
baryon-baryon correlation in the
denominator~\cite{Sato-Coalescence1981}. We expect a prominent
enhancement of the Strangeness Population Factor in a system that
passes through a deconfined partonic state, relative to what would
be observed in a system that always remained in a hadronic
phase~\cite{Cbs_PRL,Cbs_PRC}.

In this paper, a multiphase transport model (AMPT) is used to
study the effects on $S_3$ of an existing partonic phase and the
subsequent hadronic scattering. The AMPT model is suitable for
such studies, since it allows us to switch on and off the string
melting mechanism to simulate a partonic phase when the parton
density is high at early times, and it also has dynamic transport
in the early partonic phase and hadronic scattering at the late
stage. The nuclei and hypernuclei are then produced at the final
state via Wigner wave-function overlapping of their constituent
nucleons and hyperon. If the correlation present at the partonic
phase were washed out by the hadronic scattering at the later
stage, $S_3$ would have been similar, regardless of whether the
string melting mechanism is on or off in AMPT. In addition, AMPT
is also used to compute $C_{BS}$ for a comparison with $S_3$.

In a thermally equilibrated system, the yields of nuclear clusters via
the coalescence mechanism can be related to thermodynamic
quantities~\cite{harm-aos-0,BraunMunzinger:2003zd,Liu:2006my}. However,
large fluctuations away from thermal equilibrium can result in a
locally non-uniform baryon and strangeness correlation on an
event-by-event basis~\cite{Steinheimer:2008hr}. A dynamical
coalescence model has been used extensively for describing the
production of light clusters in heavy-ion
collisions~\cite{coal-model-th} at both
intermediate~\cite{inter-e-used-0,inter-e-used-2,inter-e-used-3,inter-e-used-4}
and high
energies~\cite{high-e-used-0,high-e-used-1,high-e-used-2,high-e-used-3}.
In this model, the clusters are formed in hadron phase-space at freeze
out. The probability for producing a cluster is determined by its
Wigner phase-space density without taking the binding energies into
account. The multiplicity of a $M$-hadron cluster in a heavy-ion
collision is given by
\begin{eqnarray}
N_{M} &=& G\int d\textbf{r}_{i_{1}}d\textbf{q}_{i_{1}}\cdots
d\textbf{r}_{i_{M-1}}d\textbf{q}_{i_{M-1}} \times \nonumber\\
&&
\langle\sum_{i_1>i_2>\cdots>i_M}\rho_{i}^{W}(\textbf{r}_{i_{1}},\textbf{q}_{i_{1}}\cdots\textbf{r}_{i_{M-1}},\textbf{q}_{i_{M-1}})\rangle
\label{coal-produce}
\end{eqnarray}
In Eq.~\ref{coal-produce},
$\textbf{r}_{i_{1}}$,$\cdots$,$\textbf{r}_{i_{M-1}}$ and
$\textbf{q}_{i_{1}}$,$\cdots$,$\textbf{q}_{i_{M-1}}$ are,
respectively, the $M-1$ relative coordinates and momenta in the
$M$-hadron rest frame; $\rho_{i}^{W}$ is the Wigner phase-space
density of the $M$-hadron cluster, and $\langle\cdots\rangle$
denotes the event averaging. $G$ represents the statistical factor
for the cluster; it is 1/3 for $t$, \he~\cite{g-setup} and
\hla~(\la~and \hla~have the same spin as the neutron and triton,
respectively).

To determine the Wigner phase-space densities of \he~and \hla, we take
their hadron wave functions to be those of a spherical harmonic
oscillator~\cite{inter-e-used-0,inter-e-used-2,inter-e-used-3,inter-e-used-4,harm-aos-0,harm-aos-1},

\begin{equation}
\psi =
(3/\pi^{2}b^{4})^{-3/4}\exp\left(-\frac{\rho^{2}+\lambda^{2}}{2b^{2}}\right),
\label{wave-fun}
\end{equation}
and the Wigner phase-space densities are then given by
\begin{equation}
\rho_{^{3}_{\Lambda}\rm{H}(^{3}\rm{He})}^{W}
  =
  8^{2}\exp\left(-\frac{\rho^{2}+\lambda^{2}}{b^{2}}\right)\exp\left(-(\textbf{k}^{2}_{\rho}+\textbf{k}^{2}_{\lambda})b^{2}\right)\,,
\label{wigner-density}
\end{equation}

In Eq.~\ref{wave-fun} and~\ref{wigner-density}, normal Jacobian
coordinates for a three-particle system are introduced as in
Ref.~\cite{inter-e-used-0,inter-e-used-2,inter-e-used-3,inter-e-used-4}.
($\rho$,$\lambda$) and
($\textbf{k}_{\rho}$,$\textbf{k}_{\lambda}$) are the relative
coordinates and momenta, respectively. The parameter $b$ is
determined to be 1.74 fm for
\he~\cite{inter-e-used-0,inter-e-used-2,inter-e-used-3,inter-e-used-4,AGS2004}
and 5 fm for \hla~\cite{AGS2004} from their rms radii.

\begin{figure}[htb]
\includegraphics[scale=0.43, bb=25 -45 240 280]{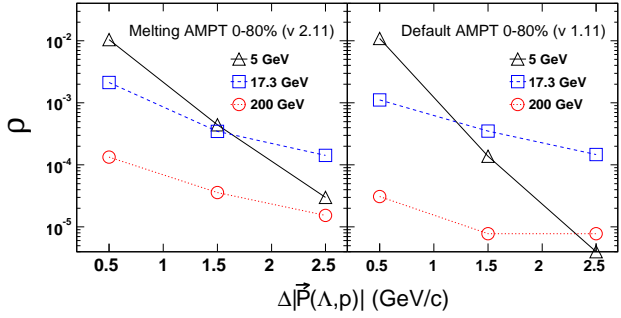}
\vspace{-0.55cm}\caption{\label{fig0}(color online) The Wigner
phase-space density $\rho$ for \hla~from melting AMPT (left panel)
and default AMPT (right panel) as a function of ($\Lambda,p$) pair
momentum. Densities are shown for \sNN= 5~GeV, 17.3~GeV and
200~GeV. The distributions have been normalized by the number of
events at each collision energy.}
\end{figure}

The coordinate and momentum space distributions of hadrons
(proton, neutron and \la) at freeze out are obtained from AMPT
model calculation~\cite{ampt}. The AMPT model has a good record of
agreement with data from RHIC~\cite{ampt}, including pion-pair
correlations~\cite{ampt-HBT} and flow~\cite{STAR-v1}. The model
has two modes: the default AMPT model (version 1.11) involves
purely hadronic interactions only, while the string melting AMPT
(version 2.11) includes a fully partonic stage at the early time
of the system evolution. Both modes have been used in the current
analysis in order to distinguish the partonic and hadronic effect.
The overlap Wigner phase-space density of the three-hadron
cluster, \hla(p,n,\la) and \he(p,p,n), is then calculated as
discussed above, and a Monte-Carlo sampling is employed to
determine if the cluster is to form a nucleus or not. A nucleus
emerges if the current sample value $\rho$ is less than
$\rho_{^{3}_{\Lambda}\rm{H}(^{3}\rm{He})}^{W}$. Figure~\ref{fig0}
depicts the \hla~Wigner phase-space density distribution as a
function of ($\Lambda,p$) pair momentum at various collision
energies in the AMPT model. The coalescence probability is larger
in melting AMPT than in the default AMPT model. The difference
increases with collision energy.

\begin{figure}[htb]
\includegraphics[scale=0.42, bb=25 -45 240 480]{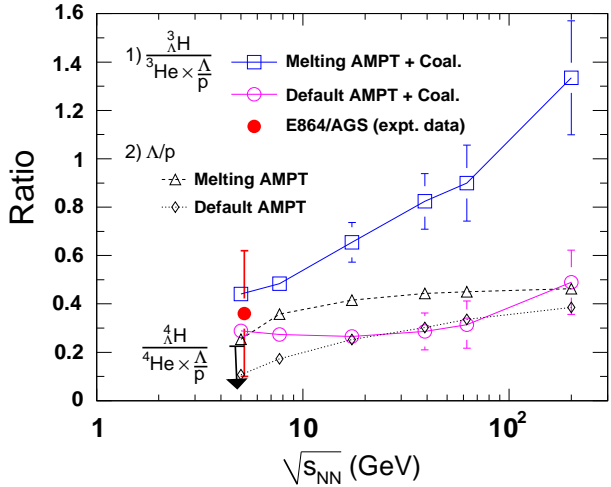}
\vspace{-0.55cm}\caption{\label{fig1}(color online) The $S_3$
ratio as a function of beam energy in minimum-bias Au+Au
collisions from default AMPT (open circles) and melting AMPT (open
squares) plus coalescence model calculations. The available data
from AGS~\cite{AGS2004} are plotted for reference. The \la$/p$
ratios from the model are also plotted.}
\end{figure}

Figure~\ref{fig1} shows the $S_3$ results for minimum-bias Au+Au
collisions at various beam energies. The definition of Strangeness
Population Factor ($S_3 =$ \hlafrac) incorporates the \la$/p$
ratio in order to remove the absolute difference in \la~and $p$
yields as a function of beam energy. It is interesting to note
that $S_3$ increases with beam energy in a system with partonic
interactions (melting AMPT) while it is almost unchanged in a
purely hadronic system (default AMPT). The measurement from
AGS~\cite{AGS2004}, in spite of large statistical uncertainty,
gives the value $\sim 1/3$. The AGS measurement of $S_4
=\mathrm{^4_\Lambda H} / ( \mathrm{^4He} \times \frac{\Lambda}{p}
)$ offers further indirect support for the lower value of $S_3$ at
the AGS~\cite{AGS2004}. A preliminary \hla/\he~result for Au+Au
collisions at 200 GeV from the STAR collaboration~\cite{RHIC09},
in combination with the measured \la/$p$ ratio from the same
experiment~\cite{PID2006,hyperon2007,PID2009}, allows us to infer
that the measured $S_3$ at RHIC is consistent with unity within
errors. These experimental results are consistent with the melting
AMPT calculations and are in contrast to the default AMPT
calculations. The data imply that the local correlation strength
between baryon number and strangeness is sensitive to the
effective number of degrees of freedom of the system created at
RHIC, and this number is significantly larger in a system
dominated by partonic interactions compared with a pure hadronic
gas. The calculated $S_3$ from melting and default AMPT modes are
close at AGS energies and are indistinguishable from the current
E864/AGS data. Moving to the top SPS energy and beyond, the
calculation using melting AMPT is more than a factor of two larger
than the results from default AMPT. It should be noted that if the
onset of deconfinement takes place at a specific beam energy, this
may result in a sharper increase of $S_3$ than the AMPT prediction
with string melting scenario. Further experimental efforts are
eagerly anticipated, including \hla~measurements as part of the
RHIC energy scan.

\begin{figure}[htb]
\includegraphics[scale=0.42, bb=20 -45 240 450]{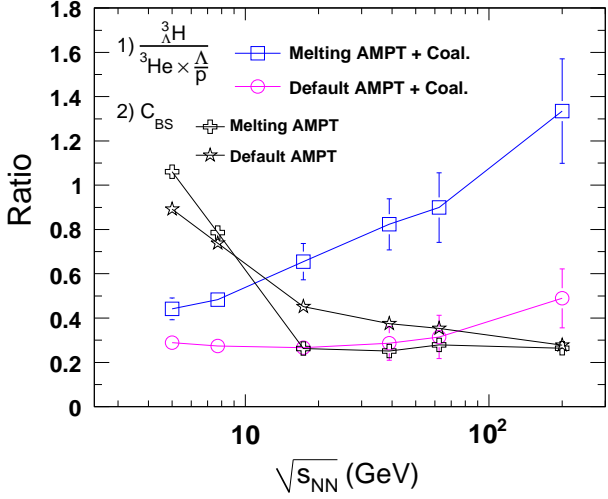}
\vspace{-0.55cm}\caption{\label{fig2}(color online) The comparison
between $S_3$ and \cbs~in minimum-bias Au+Au collisions at various
beam energies.}
\end{figure}

Furthermore, we investigate the connection between our proposed
observable $S_3$ and the original baryon-strangeness correlation
coefficient \cbs~\cite{Cbs_PRL}:
\begin{equation}
C_{BS} = -3 \frac{\langle BS \rangle - \langle B \rangle \langle S
\rangle}{\langle S^{2} \rangle - \langle S \rangle^{2}},
\label{CBS}
\end{equation}
where $B$ and $S$ are the global baryon number and strangeness in
a given rapidity window in a given event. As pointed out in
Ref.~\cite{Cbs_recombination}, a suitable rapidity window is
important to retain the fluctuation signal. We choose the rapidity
window of $-0.5 < y < 0.5$ for the present analysis.
Figure~\ref{fig2} shows the \cbs~in minimum-bias Au+Au collisions
as a function of center-of-mass energy from the AMPT model.
From top SPS to RHIC energy, the \cbs~lies between 0.2 and 0.4,
and is lower than the expected value of unity for an ideal QGP or
$\frac{2}{3}$ for a hadron gas~\cite{Cbs_PRL}. In addition, we
find that the \cbs~values from melting AMPT and default AMPT are
comparable over a wide energy range. As discussed in
Ref.~\cite{Cbs_recombination}, the recombination-like
hadronization process itself could be responsible for the
disappearance of the predicted \cbs~deconfinement signal. Detailed
study indicates that the hadronic rescattering process further
blurs the signal~\cite{Cbs_AMPT}. The \cbs~increases with an
increase of the baryon chemical potential $\mu_{B}$~\cite{Cbs_PRL}
at decreasing beam energy. The Strangeness Population Factor
$S_3$, on the other hand, increases with beam energy in a system
involving partonic interactions, as shown in Fig.~\ref{fig2}. It
carries the potential to reliably resolve the number of degrees of
freedom of the system created in heavy-ion collisions. This
suggests that the global baryon-strangeness correlation
coefficient (\cbs) is less sensitive to the local
baryon-strangeness correlation than the Strangeness Population
Factor ($S_3$) from hypernucleus production. Future precise
measurements in comparison with our calculations will provide
further insight into these physics questions that are of central
importance to relativistic heavy-ion physics.

In summary, we demonstrate that measurements of Strangeness
Population Factor $S_3$ are especially sensitive to the local
correlation strength between baryon number and strangeness, and
can serve as a viable experimental signal to search for the onset
of deconfinement in the forthcoming RHIC Beam Energy Scan.

We are grateful for discussions with Prof. H. Huang, Prof. C. M.
Ko, Prof. B. Muller, Dr. V. Koch, Dr. Z.B. Tang and H. Qiu. This
work is supported in part by the Office of Nuclear Physics, US
Department of Energy under Grants DE-AC02-98CH10886 and
DE-FG02-89ER40531, and in part by the NNSF of China under Grants
10610285, 10610286 and Chinese Academy of Science under Grants
KJCX2-YW-A14 and KJCX3-SYW-N2. Z. B. Xu is supported in part by
the PECASE Award.

\bibliography{hyperTbib}
\end{document}